\documentclass[useAMS,usenatbib,onecolumn]{mn2e}

\usepackage{dcolumn}
\usepackage{graphicx}
\usepackage{url}
\usepackage{color}
\usepackage{epstopdf}
\usepackage{textcomp}
\usepackage[T1]{fontenc}

\newcommand{\cm}{cm$^{-1}$}

\newcommand{\ai}{\textit{ab initio}}
\newcommand{\eqref}[1]{(\ref{#1})}

\newcommand{\p}{^\prime}
\newcommand{\pp}{^{\prime\prime}}

\title[ExoMol molecular line lists XII: CS]{ExoMol molecular line lists XII: Line Lists for
8 isotopologues of CS}

\date{\today}
\author[Paulose et al]{\large Geethu Paulose, Emma J. Barton,
Sergei N. Yurchenko, Jonathan Tennyson\thanks{Email: j.tennyson@ucl.ac.uk (JT)} \\
Department of Physics and Astronomy, University College London, London WC1E 6BT, UK}
\date{Accepted XXXX. Received XXXX; in original form XXXX}

\pagerange{\pageref{firstpage}--\pageref{lastpage}} \pubyear{2015}

\begin{document}

\maketitle

\begin{abstract}

  Comprehensive vibration-rotation line lists for eight isotopologues
  of carbon monosulphide (CS) ($^{12}$C$^{32}$S, $^{12}$C$^{33}$S,
  $^{12}$C$^{34}$S, $^{12}$C$^{36}$S, $^{13}$C$^{32}$S,
  $^{13}$C$^{33}$S, $^{13}$C$^{34}$S, $^{13}$C$^{36}$S) in their
  ground electronic states are calculated. These line lists are
  suitable for temperatures up to 3000 K. A spectroscopically-determined potential
  energy curve (PEC) and dipole moment curve (DMC) are taken from
  literature. This PEC is
  adapted to suit our method prior to the computation of ro-vibrational energies.
 The calculated energies are then substituted by
  experimental energies, where available, to improve the accuracy of the line lists.
  The {\it ab initio} DMC is used without refinement to
  generate Einstein A coefficients. Full line lists of
  vibration-rotation transitions and partition functions are made
  available in an electronic form as supporting information to this
  paper and at \url{www.exomol.com}.

\end{abstract}

\textit{molecular data; opacity; astronomical data bases: miscellaneous; planets and satellites: atmospheres; stars: low-mass}

\label{firstpage}

\section{Introduction}

It was the development of radio astronomy that led to the realisation
that the majority of our galaxy, and hence the universe, is dominated
by molecular processes. More than 150 molecules have been detected so
far in the interstellar medium (ISM) by direct observation of their
spectra. Carbon monosulphide (CS) is one of these molecules
\citep{71PeSoWi.CS} and is a diatomic of both atmospheric as well as
astrophysical interest.  In the Earth's atmosphere, CS plays a role in
the formation of aerosols, in particular carbonyl sulphide (OCS) in
the troposphere \citep{13LiDeZh.CS}.

In the solar system CS has been observed in comets \citep{07CaAlBo.CS}
and the collision of comet Shoemaker-Levy 9 \citep{jt154} led to its
detection in the atmosphere of Jupiter \citep{03MoMaMa.CS}.
Astronomically the molecule has been observed in a variety of objects
such as carbon-rich stars \citep{78BrGoSt.CS,85BoSexx.CS,06AgCexx.CS},
star forming regions \citep{13DaBaCr.CS} and dense interstellar clouds
\citep{00NiBeHj.CS,02McSiLa.CS,15ScShWa.CS}. In fact CS is one of the
most abundant sulphur-containing species in interstellar clouds
\citep{11ShLiZh.CS,15BiBaxx.CS} with  several isotopologues long
detected outside the Milky Way
\citep{85HeBaxx.CS,89MaHeWia.CS,89MaHeWib.CS,93HeMaWi.CS}.

The numerous astronomical detections of CS and the importance of the
molecule in our own atmosphere has motivated copious laboratory studies.
Experimentally the CS spectrum has been studied in wavelength regions
ranging from the microwave to the ultraviolet (UV). A pioneering
study was carried out by \citet{34CaShxx.CS} whom discovered the main
$A~^{1}\Pi$ - $X~^{1}\Sigma^{+}$ transition in the visible.
Additional electronic transitions in the visible to near UV have been
investigated by, for example, \citet{72BeNgSu.CS}, \citet{77CoHoRo.CS}
and \citet{87StYoSm.CS}.

\citet{55MoBixx.CS} made the first measurements of rotational lines for the
$^{12}$C$^{32}$S, $^{12}$C$^{33}$S, $^{12}$C$^{34}$S and $^{13}$C$^{32}$S
isotopomers in the microwave region. Transitions in this region have also been
observed by, for example, \citet{74LoKrxx.CS}.
Early work in the millimetre wave region began with measurements of
rotational $^{12}$C$^{32}$S and $^{12}$C$^{34}$S lines by
\citet{63KeSaWi.CS}, later extended by \citet{82BoDeDe.CS} who
additionally observed $^{13}$C$^{32}$S.  \citet{81BoDeDe.CS} also
presented a study of the millimetre spectrum of rarer isotopologues
$^{12}$C$^{33}$S,$^{12}$C$^{36}$S, $^{13}$C$^{33}$S and
$^{13}$C$^{34}$S. More recent work in the region has been carried out
by \citet{99AhWixx.CS} ($^{12}$C$^{32}$S, $^{12}$C$^{34}$S,
$^{12}$C$^{33}$S, $^{13}$C$^{32}$S, $^{12}$C$^{36}$S,
$^{13}$C$^{33}$S, $^{13}$C$^{34}$S), \citet{03KiYaxx.CS}
($^{12}$C$^{32}$S, $^{12}$C$^{34}$S) and \citet{03GoMyTh.CS}
($^{12}$C$^{32}$S, $^{12}$C$^{34}$S, $^{12}$C$^{33}$S,
$^{13}$C$^{32}$S).

The first study in the infra-red region was performed by
\citet{77Toxxxx.CS} who measured the $v = 2-0$ vibrational band of the
main isotope while \citet{79ToOlxx.CS} and \citet{79YaHixx.CS}
measured several $\Delta v$ = 1 bands of $^{12}$C$^{32}$S,
$^{12}$C$^{33}$S, $^{12}$C$^{34}$S and $^{13}$C$^{32}$S. The resulting
molecular parameters were later refined by \citet{84WiDaPe.CS} and
\citet{87BuLoHa.CS}.  \citet{84WiDaPe.CS} measured many $\Delta v$ = 2
bands for vibrational levels up to $v$ = 8 while \citet{87BuLoHa.CS}
obtained high resolution measurements of the 1-0 band, and 2-1 band
for the main isotope, of $^{12}$C$^{33}$S,
$^{12}$C$^{34}$S and $^{13}$C$^{32}$S for $J$ up to 41, 28, 32 and 28
respectively. $\Delta v$ = 1 bands up to $v$ = 9-8 for the main isotope
were later measured by \citet{95RaBeDa.CS}.
The most recent research on CS infra-red spectra has
been performed by \citet{15UeHoSa.CS}, who reported $\Delta v$ = 1 transitions
of $^{13}$C$^{32}$S up to $v$ = 5-4. Additionally they measured $\Delta v$ = 1
transitions of $^{12}$C$^{32}$S up to $v$ = 7-6 to a higher accuracy than
\citet{95RaBeDa.CS}.

Using the vibration-rotation and pure rotation data on this molecule
available to them, \citet{92CoHaxx.CS} derived a spectroscopic
potential energy curve (PEC) that reproduced all the input
experimental data within experimental error. This PEC is the starting
point for the current work.

Transition probabilities or Einstein A coefficients have been provided by
\citet{85BoSexx.CS} and \citet{87PiTiCh.CS}. The latter produced transition
lists for  rotational quantum numbers $J\p - J = \pm 1$ for $J \leq 200$, and
vibrational $v \leq 20$, though only for the four most abundant isotopes
($^{12}$C$^{32}$S, $^{12}$C$^{33}$S, $^{12}$C$^{34}$S and $^{13}$C$^{32}$S).
Line lists for all isotopologues of CS, including some
rotation-vibration transitions, are available from CDMS \citep{cdms}. These
were constructed from experimental data obtained by \citet{82BoDeDe.CS},
\citet{81BoDeDe.CS}, \citet{99AhWixx.CS}, \citet{03KiYaxx.CS} ,
\citet{03GoMyTh.CS}, \citet{87BuLoHa.CS}, \citet{95RaBeDa.CS},
\citet{84WiDaPe.CS} and the only available experimental measurements of the
dipole moment \citep{68WiCoxx.CS}. The line lists are very accurate and
recommended for use in radio astronomy, though are limited to $v \leq 4$ and
$J \leq 99$ and hence temperatures, $T$ below about 500~K. The aim of this
work is to produce comprehensive line lists for all stable isotopologues of
CS suitable for modelling hot environments, such as carbon stars ($T \sim
3000$ K).

The ExoMol project aims to provide line lists on all the molecular
transitions of importance in the atmospheres of planets \citep{jt528}.
The ExoMol methology has already been applied to a number of diatomic
molecules: BeH, MgH and CaH \citep{jt529}, SiO \citep{jt563}, NaCl and
KCl \citep{jt583}, PN \citep{jt590}, AlO \citep{jt598} and NaH
\citep{jt605}. In this paper, we present ro-vibrational transition
lists and associated spectra for all stable isotopologues of CS.

\section{Method}

The line lists for all eight isotopologues of CS, which we have named
JnK, were obtained by solving the Schr\"{o}dinger equation allowing for
Born-Oppenheimer Breakdown (BOB) effects using the program LEVEL8.0
\citep{lr07}. In principle the calculations were initiated using the
spectroscopic PEC of \citet{92CoHaxx.CS}. In practice, as described
below, the PEC was first expressed in a form compatible with LEVEL.
The PEC was then adapted to improve the results of calculations
performed using LEVEL. After computation the line lists are improved by
replacing calculated energies with experimentally derived energies, where
available, and shifting the remaining energies to maintain LEVEL predicted
energy level separations. A theoretical dipole moment curve (DMC) from
\citet{87PiTiCh.CS} was also employed.

\subsection{Potential Energy Curve}

We did not generate a new PEC for CS. A full set of potential parameters
representing a very accurate PEC is already available from
\citet{92CoHaxx.CS}. The authors employed data for four isotopomers
($^{12}$C$^{32}$S, $^{12}$C$^{33}$S, $^{12}$C$^{34}$S and $^{13}$C$^{32}$S)
in a least-squares fit to a potential function in the Born-Oppenheimer
approximation and BOB functions to determine a PEC valid for all
isotopologues. They set the dissociation energy $D_{e}$ to 59300.0 \cm\ and
expressed their fitted potential $U_{\rm CS}^{\rm BO}(R)$ as a Morse
potential with a  variable $\beta$:
\begin{equation}
U_{\rm CS}^{\rm BO}(R) = D_{\rm e} \left(1 - \exp\left[-\beta (R)(R - R_{\rm e})\right]\right)^{2},
\end{equation}
where:
\begin{equation}
\beta (R) = \beta_{0} + \beta_{1}(R - R_{\rm e}) + \beta_{2}(R - R_{\rm e})^{2} +...
\end{equation}
The isotopically invariant breakdown functions were modelled as:
\begin{equation}
U_{\rm C}(R) = \Sigma_{i=1} u_{i}^{\rm C}(R - R_{\rm e})^{i},
\end{equation}
and
\begin{equation}
U_{\rm S}(R) = \Sigma_{i=1} u_{i}^{\rm S}(R - R_{\rm e})^{i},
\end{equation}
while the $J$-dependent non-adiatbatic breakdown function was modelled as:
\begin{equation}
q_{\rm  CS}(R) = q^{\rm C}(R)/M_{\rm C} + q^{\rm S}(R)/M_{\rm S},
\end{equation}
where:
\begin{equation}
q^{(k)}(R) = \Sigma_{i=1} \rho_{i}^{(k)}(R - R_{\rm e})^{i},
\end{equation}
such that $q^{(k)}(R_{\rm e}) = 0$. These were applied in the effective
radial Hamiltonian according to:
\begin{equation}
H_{\rm CS}(R) = -\beta_{at}^{2} \nabla_{R}^{2} + U_{\rm CS}^{eff}(R) + (\beta_{at}^{2}/R^{2})J(J+1)[1+q_{\rm CS}(R)],
\end{equation}
where $\beta_{at}^{2} = \hbar^{2}/2\mu_{at}$, defined with the atomic masses and:
\begin{equation}
U_{\rm CS}^{eff}(R) = U_{\rm CS}^{\rm BO}(R) + U_{\rm C}(R)/M_{\rm C} + U_{\rm S}(R)/M_{\rm S},
\end{equation}
%\citet{92CoHaxx.CS}.
The variable $\beta$ Morse is not implemented in LEVEL8.0. The
coefficients from \citet{92CoHaxx.CS}, given in Column II of
Table~\ref{tab:potpar}, were hence used to generate turning points for a range of
internuclear distances. The turning points were used directly in LEVEL.

Employing the potential parameters of \citet{92CoHaxx.CS} in this way we
could not reproduce the vibrational energies to the spectroscopic accuracy
achieved by \citet{92CoHaxx.CS}. By applying small `corrections' to potential
parameters $u_{1}^{\rm C}$, $u_{1}^{\rm S}$, $\rho_{1}^{\rm S}$ and
$\rho_{2}^{\rm S}$, we were able to predict the ro-vibrational energies up to
$v$ = 9, and experimental frequencies, for $^{12}$C$^{32}$S and $^{13}$C$^{32}$S to within 0.02 \cm\
and 0.04 \cm\ respectively
(see JnK columns in Table~\ref{tab:Epar} and Table~\ref{tab:freq}). To give an example, the residual
(obs-calc) for $^{12}$C$^{32}$S $T_1$ without the `corrections' was 0.027 \cm\ and 0.007 \cm\
after correction. The term `correction' is used tentatively in this context as, although the
modification of the potential parameters improved the present results, this
is not an improvement on the variable $\beta$ Morse presented by
\citet{92CoHaxx.CS}. The potential parameters used in this work are given as
Column III of Table~\ref{tab:potpar}.

To further improve our results we
took advantage of the ExoMol format used to store the line list, see Section
3. Put simply this is a states file containing level energies and a
transitions file detailing allowed energy level couplings. The advantage of
the format is it gives the option of replacing calculated energies with more
refined or experimental energies such that, when the files are unpacked to
produce the line list, more accurate line frequencies are computed, see \cite{jt570} for example.

First we attempted to refine our ro-vibrational energies ($E_{v,J}^{\rm JnK}$) for all
isotopologues using the vibrational ($J$ = 0, $v \leq $20) energies given in
\citet{92CoHaxx.CS} ($E_{v,0}^{\rm Cox}$) and the formula:
\begin{equation}
\label{eq:cox1}
E_{v,J}^{{\rm JnK}-\rm{Cox}} = \left(E_{v,J}^{\rm JnK} - E_{v,0}^{\rm JnK}\right) + E_{v,0}^{\rm Cox}
\end{equation}
Ro-vibrational energies for $v$ > 20 were shifted to maintain the energy level separations
predicted by LEVEL according to:
\begin{equation}
\label{eq:cox2}
E_{v,J}^{{\rm JnK}-{\rm Cox}} = \left(E_{v,J}^{\rm JnK} - E_{v-1,J}^{\rm JnK}\right) + E_{v-1,J}^{\rm Cox}
\end{equation}
We could then reproduce the
vibrational energies to the same spectroscopic accuracy achieved by \citet{92CoHaxx.CS},
however not all the line frequency predictions improved
(see JnK-Cox columns in Table~\ref{tab:freq}).

This is likely due to the fact experimental data available to \citet{92CoHaxx.CS} was limited
to $J \leq$41 and $J \leq$28 while \citet{95RaBeDa.CS} and \citet{15UeHoSa.CS} assigned lines for
$J$ up to 113 and 86 for $^{12}$C$^{32}$S and $^{13}$C$^{32}$S respectively.

Therefore we decided to determine experimental energies directly from frequencies
measured by \citet{95RaBeDa.CS} and \citet{15UeHoSa.CS} using the
measured active rotation-vibration energy levels (MARVEL) technique
\citep{jt412} which involves inverting transition frequencies
to extract experimental level energies.
\citet{15UeHoSa.CS} is the more accurate experimental study and thence energies extracted
from these frequencies were used preferentially over those extracted from \citet{95RaBeDa.CS}
frequencies where possible.
We extracted 733 energies in total for the main isotopologue and 341 energies for $^{13}$C$^{32}$S,
see Table~\ref{tab:extracted}.

$^{12}$C$^{32}$S and $^{13}$C$^{32}$S energies for the experimental ranges
were replaced with experimentally derived energies. Ro-vibrational energies for ($v$, $J$)
outside the experimental ranges were shifted to maintain the energy level separations predicted by
LEVEL according to the following equations.
For $v$ < $v^{\rm Exp}_{\rm max}$ and $J$ > $J^{\rm Exp}_{\rm max}$:
\begin{equation}
\label{eq:exp1}
E_{v,J}^{{\rm JnK}-{\rm Exp}} = \left(E_{v,J}^{\rm JnK} - E_{v,0}^{\rm JnK}\right) + E_{v,0}^{\rm Exp}
\end{equation}
For $v$ > $v^{\rm Exp}_{\rm max}$:
\begin{equation}
\label{eq:exp2}
E_{v,J}^{{\rm JnK}-{\rm Exp}} = \left(E_{v,J}^{\rm JnK} - E_{v-1,J}^{\rm JnK}\right) + E_{v-1,J}^{{\rm JnK}-{\rm Exp}}
\end{equation}

The experimental frequencies for $^{12}$C$^{32}$S and $^{13}$C$^{32}$S, by default, were reproduced
almost exactly using this method. The largest residual is < 0.001 \cm. Hence the accuracy of these
line lists should be equal to the experimental accuracies, which are expected to be 0.012 \cm\
and 0.01 \cm\ for \citet{95RaBeDa.CS} and \citet{15UeHoSa.CS} respectively.

For $^{12}$C$^{33}$S and $^{12}$C$^{34}$S \citet{87BuLoHa.CS} measured infra-red $v$ = 1 -- 0 absorption frequencies
at 0.004 \cm unapodized resolution. The Coxon refined energies, see Eqs.~\ref{eq:cox1} and \ref{eq:cox2},
reproduce these frequencies to very high precision (see Table~\ref{tab:other1}),
as would be expected considering
they fitted to these frequencies. The Coxon refined energies also represent an improvement on the unrefined JnK energies
(see Table~\ref{tab:other1}), therefore we chose to employ them in our final line lists for these two isotopologues.

For the remaining isotopologues, with the exception of $^{13}$C$^{36}$S
which has not been observed experimentally, we have only measurements of rotational
frequencies from \citet{99AhWixx.CS} to compare with. These are expected to have
an accuracy of at least 0.00002 \cm. As can been seen in Table~\ref{tab:other2} the agreement is excellent.
Due to our methods, unrefined JnK and Coxon refined energies predict the same rotational frequencies.
However, since the Coxon refined energies improved ro-vibrational frequency predictions for other isotopologues,
these are employed in our final line lists for the remaining four isotopologues.

An overview of the energy level content of our final 'hybrid' line lists is given in Table ~\ref{tab:hybrid}.
Although we have used terms JnK, JnK-Cox and JnK-Exp in the text to refer to unrefined, Coxon refined and
experimentally substituted energies respectively, the final line lists as provided in supplementary data and on
www.exomol.com are simply named JnK.

\begin{table}
\caption{Coefficients of the Born-Oppenheimer potential and radial functions
for the $X~^{1}\Sigma^{+}$ state of CS.}
\label{tab:potpar} \footnotesize
\begin{center}
\begin{tabular}{lll}
\hline
Parameter & \citet{92CoHaxx.CS}$^{1}$ & This Work \\
\hline
$R_{\rm e}$ \AA	&	1.5348175 (27)	&	1.5348175 	\\
$\beta_{0}$	&	1.89830910 (907)	&	1.89830910 	\\
$\beta_{1}$	&	0.0173292 (2255)	&	0.0173292 	\\
$\beta_{2}$	&	0.1220304 (5245)	&	0.1220304 	\\
$\beta_{3}$	&	0.0317616 (6808)	&	0.0317616	\\
$\beta_{4}$	&	0.0437801 (7351)	&	0.0437801 	\\
$\beta_{5}$	&	0.028153 (1245)	&	0.028153 	\\
$\mathbf{u_{1}^{\rm C}}$	&	\textbf{-473.55 (11.10)}	&	\textbf{-474}	\\
$u_{2}^{\rm C}$	&	1631.08 (21.83)	&	1631.08	\\
$u_{3}^{\rm C}$	&	-1380.3 (640.8)	&	-1380.3 	\\
$u_{4}^{\rm C}$	&	-13714.9 (730.8)	&	-13714.9 	\\
$\mathbf{u_{1}^{\rm S}}$	&	\textbf{-448.24 (18.61)}	&	\textbf{-448}	\\
$u_{2}^{\rm S}$	&	1124.90 (52.07)	&	1124.90 	\\
$u_{3}^{\rm S}$	&	-1245.2 (116.2)	&	-1245.2	\\
$\mathbf{\rho_{1}^{\rm S}}$	&	\textbf{-0.0041274 (3021)}	&	\textbf{-0.0038353}	\\
$\mathbf{\rho_{2}^{\rm S}}$	&	\textbf{-0.0008036 (4672)}	&	\textbf{-0.0003364}	\\
\hline
\end{tabular}

\noindent
$^{1}$Uncertainties are given in parentheses in units of the last digit.

\end{center}
\end{table}

\begin{table}
\caption{A comparison of theoretically and experimentally derived vibrational
term values for $^{12}$C$^{32}$S and $^{13}$C$^{32}$S in \cm.}
\label{tab:Epar} \footnotesize
\begin{center}
\begin{tabular}{llrrrr}
\hline
$T_{v}$ & Experiment & Calculated & Obs-Calc & \citet{92CoHaxx.CS} & \\
& & This Work & This Work & & \\
& (JnK-Exp) & (JnK) & (JnK) & (JnK-Cox) & (JnK-Cox) \\
\hline
 & & $^{12}$C$^{32}$S & & & \\
\citet{15UeHoSa.CS} & & & & & \\
$T_{1}$	&	1272.162085	&	1272.1690	&	-0.0069	&	1272.16214	&	-0.00006	\\
$T_{2}$	&	2531.353715	&	2531.3486	&	0.0051	&	2531.35384	&	-0.00013	\\
$T_{3}$	&	3777.597715	&	3777.5899	&	0.0078	&	3777.59800	&	-0.00029	\\
$T_{4}$	&	5010.916850	&	5010.9087	&	0.0082	&	5010.91739	&	-0.00054	\\
$T_{5}$	&	6231.333760	&	6231.3229	&	0.0109	&	6231.33456	&	-0.00080	\\
$T_{6}$	&	7438.870820	&	7438.8584	&	0.0124	&	7438.87183	&	-0.00101	\\
$T_{7}$	&	8633.550080	&	8633.5338	&	0.0163	&	8633.55125	&	-0.00117	\\
\citet{95RaBeDa.CS} & & & & & \\
$T_{8}$ & 9815.39254    &    9815.3752	  &    0.0173   &   9815.39457  &	-0.0020 \\
$T_{9}$ & 10984.42006   &   10984.4025    &   0.0176    &  10984.42297  & -0.0029	\\
 & & $^{13}$C$^{32}$S & & & \\
\citet{15UeHoSa.CS} & & & & & \\
$T_{1}$	&	1236.315929	&	1236.3524	  &   -0.0364  &  1236.31591    	 &	 0.00002     \\
$T_{2}$	&	2460.388235	&	2460.3782         &   0.0100   &  2460.39043    	 &	 -0.00220      \\
$T_{3}$	&	3672.237852	&	3672.2234	& 0.0145    &  3672.24479    	 &	 -0.00694      \\
$T_{4}$	&	4871.88567	&	4871.8831	&	  0.0026    &      -	         &	-       \\
$T_{5}$	&	6059.35246	&	6059.3547	&	  -0.0021    &      -	         &	-       \\
\hline
\end{tabular}
\end{center}
\end{table}

\begin{table}
\caption{Comparison of predicted ro-vibrational frequencies, in \cm, with experimental line
positions measured by \citet{95RaBeDa.CS} and \citet{15UeHoSa.CS} for $^{12}$C$^{32}$S and $^{13}$C$^{32}$S.}
\label{tab:freq}
\begin{center}
\begin{tabular}{lllllllll}
\hline
$J^{\prime}$ & $J^{\prime\prime}$ & $v^{\prime}$ & $v^{\prime\prime}$ & Experimental & Calculated & Obs-Calc & Calculated & Obs-Calc \\
 & & & & JnK-Exp & JnK & JnK & JnK-Cox & JnK-Cox \\
\hline
$^{12}$C$^{32}$S & & & & & & & & \\
& & & & \citet{15UeHoSa.CS} & & & & \\
10	& 	9	& 	1	& 	0	& 	1287.847083	& 	1287.8541	& 	-0.0070	& 	1287.8472	& 	-0.0001 \\
10	& 	11	& 	1	& 	0	& 	1253.542068	& 	1253.5490	& 	-0.0070	& 	1253.5421	& 	-0.0000	\\
9	& 	10	& 	2	& 	1	& 	1242.440398	& 	1242.4286	& 	0.0118	& 	1242.4407	& 	-0.0003  \\
11	& 	10	& 	2	& 	1	& 	1276.248067	& 	1276.2363	& 	0.0118	& 	1276.2484	& 	-0.0003	\\
100	& 	101	& 	2	& 	1	& 	1040.852470	& 	1040.8485	& 	0.0039	& 	1040.8606	& 	-0.0082	\\
20	& 	21	& 	6	& 	5	& 	1172.020927	& 	1172.0201	& 	0.0008	& 	1172.0219	& 	-0.0010	\\
21	&	20	&	6	&	5	&	1237.819388	& 	1237.8182	& 	0.0012	& 	1237.8200	& 	-0.0006	\\
78	&	79	&	6	&	5	&	1049.137636	& 	1049.1376	& 	0.0000	& 	1049.1394	& 	-0.0018	\\
10	&	11	&	7	&	6	&	1176.836843	& 	1176.8360	& 	0.0008	& 	1176.8400	& 	-0.0032	\\
15	&	14	&	7	&	6	&	1216.680436	& 	1216.6786	& 	0.0018	& 	1216.6827	& 	-0.0022	\\
63	&	64	&	7	&	6	&	1072.087367	& 	1072.0888	& 	-0.001	& 	1072.0928	& 	-0.0054	\\
& & & & \citet{95RaBeDa.CS} & & & &  \\
102	& 	103	& 	1	& 	0	& 	1047.2868	& 	1047.2910	& 	-0.004	&	1047.2841	&	0.0027	\\
107	& 	106	& 	1	& 	0	& 	1371.8550	& 	1371.8529	& 	0.0021	& 	1371.8459	& 	0.0090	\\
107	& 	106	& 	2	& 	1	& 	1357.5752	& 	1357.5763	& 	-0.0011	& 	1357.5884	& 	-0.0132	\\
89	&	88	& 	6	&	5	&	1296.2749	&	1296.2749	&	0.0000	&	1296.2767	&	-0.0018	\\
89	&	88	&	7	&	6	&	1282.3246	&	1282.3263	&	-0.0017	&	1282.3303	&	-0.0057	\\
25	&	26	&	8	&	7	&	1137.7464	&	1137.7446	&	0.0018	&	1137.7465	&	-0.0001	\\
30	&	31	&	8	&	7	&	1128.3927	&	1128.3910	&	0.0017	&	1128.3928	&	-0.0001	\\
52	&	53	&	8	&	7	&	1084.0478	&	1084.0496	&	-0.0018	&	1084.0514	&	-0.0036	\\
59	&	58	&	8	&	7	&	1251.2061	&	1251.2079	&	-0.0018	&	1251.2098	&	-0.0037	\\
25	&	26	&	9	&	8	&	1125.2391	&	1125.2367	&	0.0024	&	1125.2379	&	0.0012	\\
28	&	27	&	9	&	8	&	1207.1842	&	1207.1833	&	0.0009	&	1207.1844	&	-0.0002	\\
52	&	53	&	9	&	8	&	1071.8531	&	1071.8548	&	-0.0017	&	1071.8559	&	-0.0028	\\
59	&	58	&	9	&	8	&	1237.6787	&	1237.6766	&	0.0021	&	1237.6778	&	0.0009	\\
$^{13}$C$^{32}$S & & & & & & &  & \\
& & & & \citet{15UeHoSa.CS} & & & & \\
2	&	1	&	1	&	0	&	1239.369011	&	1239.4051	&	-0.0360	&	1239.3686	&	0.0004	\\
79	&	80	&	1	&	0	&	1080.972174	&	1080.9775	&	-0.0053	&	1080.9411	&	0.0311	\\
12	&	13	&	2	&	1	&	1203.322346	&	1203.2753	&	0.0470	&	1203.3239	&	-0.0015	\\
70	&	71	&	2	&	1	&	1089.992782	&	1089.9459	&	0.0469	&	1089.9944	&	-0.0016	\\
7	&	8	&	3	&	2	&	1199.383201	&	1199.3761	&	0.0071	&	1199.3855	&	-0.0023	\\
53	&	52	&	3	&	2	&	1276.188741	&	1276.1977	&	-0.0089	&	1276.2070	&	-0.0182	\\
6	&	7	&	4	&	3	&	1188.850787	&	1188.8625	&	-0.0117	&	1188.8411	&	0.0097	\\
64	&	65	&	4	&	3	&	1080.169544	&	1080.1756	&	-0.0061	&	1080.1542	&	0.0154	\\
14	&	13	&	5	&	4	&	1207.300671	&	1207.3057	&	-0.0051	&	1207.3057	&	-0.0051	\\
45	&	46	&	5	&	4	&	1107.707636	&	1107.7159	&	-0.0083	&	1107.7159	&	-0.0083	\\
\hline
\end{tabular}
\end{center}
\end{table}

\begin{table}
\caption{Comparison of predicted ro-vibrational frequencies, in \cm, with experimental line
positions measured by \citet{87BuLoHa.CS} for $^{12}$C$^{33}$S and $^{12}$C$^{34}$S.}
\label{tab:other1}
\begin{center}
\begin{tabular}{lllllllll}
\hline
$J^{\prime}$ & $J^{\prime\prime}$ & $v^{\prime}$ & $v^{\prime\prime}$ & Experimental & Calculated & Obs-Calc & Calculated & Obs-Calc \\
 & & & &  & JnK & JnK & JnK-Cox & JnK-Cox \\
\hline
$^{12}$C$^{33}$S & & & & & & & & \\
3	&	2	&	1	&	0	&	1271.73537	&	1271.736018	&	-0.000648	&	1271.735366	&	 0.000004	\\
4	&	5	&	1	&	0	&	1258.72386	&	1258.724631	&	-0.000771	&	1258.723979	&	-0.000119	\\
29	&	28	&	1	&	0	&	1308.72699	&	1308.728068	&	-0.001078	&	1308.727416	&	-0.000426	\\
25	&	26	&	1	&	0	&	1221.09717	&	1221.098153	&	-0.000983	&	1221.097501	&	-0.000331	\\
$^{12}$C$^{34}$S & & & & & & & & \\
3	&	2	&	1	&	0	&	1266.78054	&	1266.775300	&	0.005240    &	1266.780744	&	-0.000204	\\
4	&	5	&	1	&	0	&	1253.87137	&	1253.865751	&	0.005619	&	1253.871195	&	0.000175	\\
35	&	34	&	1	&	0	&	1310.80236	&	1310.797693	&	0.004667	&	1310.803137	&	-0.000777	\\
35	&	36	&	1	&	0	&	1197.09619	&	1197.091827	&	0.004363	&	1197.097271	&	-0.001081	\\
\hline
\end{tabular}
\end{center}
\end{table}

\begin{table}
\caption{Comparison of predicted rotational frequencies, in \cm, with experimental line
positions measured by \citet{99AhWixx.CS} for $^{12}$C$^{36}$S, $^{13}$C$^{33}$S and $^{13}$C$^{34}$S.}
\label{tab:other2}
\begin{center}
\begin{tabular}{lllllll}
\hline
$J^{\prime}$ & $J^{\prime\prime}$ & $v^{\prime}$ & $v^{\prime\prime}$ & Experimental & Calculated & Obs-Calc \\
 & & & &  & JnK \ JnK-Cox & JnK \ JnK-Cox \\
\hline
$^{12}$C$^{36}$S & & & & & & \\
6	&	5	&	0	&	0	&	9.507279	&	9.507297	&	-0.000018	\\
22	&	21	&	1	&	1	&	34.561685	&	34.561709	&	-0.000024	\\
$^{13}$C$^{33}$S & & & & & & \\
6	&	5	&	0	&	0	&	9.173973	&	9.173971	&	0.000002	\\
20	&	19	&	0	&	0	&	30.545844	&	30.545846	&	-0.000002	\\
$^{13}$C$^{34}$S & & & & & & \\
20	&	19	&	0	&	0	&	30.293201	&	30.293299	&	-0.000098	\\
13	&	12	&	2	&	2	&	19.429173	&	19.429154	&	0.000019	\\
\hline
\end{tabular}
\end{center}
\end{table}

\begin{table}
\caption{Summary of Energies Extracted from Experimental Frequencies.}
\label{tab:extracted} \footnotesize
\begin{center}
\begin{tabular}{lllll}
\hline
v & Jmax & Total Extracted & Using \citet{15UeHoSa.CS} & Using \citet{95RaBeDa.CS} \\
\hline
$^{12}$C$^{32}$S & & & &\\
0 & 106 & 107 & 87 & 20 \\
1 & 106 & 107 & 102 & 5 \\
2 & 101 & 102 & 94 & 8 \\
3 & 93 & 94 & 86 & 8 \\
4 & 89 & 90 & 89 & 1 \\
5 & 89 & 80 & 62 & 18 \\
6 & 76 & 55 & 41 & 14 \\
7 & 70 & 46 & 24 & 22 \\
8 & 59 & 32 & 0 & 32 \\
9 & 59 & 32 & 0 & 32 \\
$^{13}$C$^{32}$S & & & & \\
0 & 80 & 78 & 78 & 0 \\
1 & 71 & 70 & 70 & 0 \\
2 & 70 & 69 & 69 & 0 \\
3 & 65 & 64 & 64 & 0 \\
4 & 46 & 30 & 30 & 0 \\
5 & 46 & 30 & 30 & 0 \\
\hline
\end{tabular}
\end{center}
\end{table}

\begin{table}
\caption{Overview of sources of energy levels used in the JnK 'hybrid' line lists.}
\label{tab:hybrid} \footnotesize
\begin{center}
\begin{tabular}{llll}
\hline
Isotopologue & Experimental Energies & Coxon Energies & Shifted Energies \\
\hline
$^{12}$C$^{32}$S & v $\leq$ 9, J $\leq$ 106 & None & v > 9, J > 106 using Eq.~\ref{eq:exp1} and Eq.~\ref{eq:exp2} \\
$^{12}$C$^{33}$S & None & v $\leq$ 7 using Eq.~\ref{eq:cox1} & v > 7 using Eq.~\ref{eq:cox2} \\
$^{12}$C$^{34}$S & None & v $\leq$ 7 using Eq.~\ref{eq:cox1} & v > 7 using Eq.~\ref{eq:cox2} \\
$^{12}$C$^{36}$S & None & v $\leq$ 7 using Eq.~\ref{eq:cox1} & v > 7 using Eq.~\ref{eq:cox2} \\
$^{13}$C$^{32}$S & v $\leq$ 5, J $\leq$ 80 & None & v > 5, J > 80 using Eq.~\ref{eq:exp1} and Eq.~\ref{eq:exp2} \\
$^{13}$C$^{33}$S & None & v $\leq$ 3 using Eq.~\ref{eq:cox1} & v > 3 using Eq.~\ref{eq:cox2} \\
$^{13}$C$^{34}$S & None & v $\leq$ 3 using Eq.~\ref{eq:cox1} & v > 3 using Eq.~\ref{eq:cox2}  \\
$^{13}$C$^{36}$S & None & v $\leq$ 3 using Eq.~\ref{eq:cox1} & v > 3 using Eq.~\ref{eq:cox2}  \\
\hline
\end{tabular}
\end{center}
\end{table}

We note that the behaviour of the potential curve for longer internuclear distances than 3
\AA\ was found to be divergent and non-physical resulting in the molecule not
dissociating properly. This is consistent with the findings of
\citet{97CoCoxx.diatom}, that the model employed by \citet{92CoHaxx.CS} does
not account for the inverse-power behaviour of the PEC at long range. For
this reason we only generated turning-points for internuclear distances
between $R_{\rm min} = 1.00$ \AA\ and $R_{\rm max} = 3.00$ \AA\ with a grid
spacing of $0.0007$ \AA. This has consequences for the temperature range
considered. Based on our partition sum, see Section 2.3, this range now
extends to 3000 K.

One approach proposed by \citet{00HaLexx.diatom} to overcome the problem
is to employ a Modified Lennard-Jones (MLJ) potential function in
place of the Morse variable $\beta$ \citep{97CoCoxx.diatom}. As it is
possible to produce a high temperature line list without this
treatment, it is beyond the scope of this work.

\subsection{Dipole Moment Curve}

The only experimental dipole moment data for CS in the literature are Stark
measurements of the $J$ = 1 $\leftarrow$ 0 matrix elements for $v$ = 0 and
$v$ = 1 \citep{68WiCoxx.CS}. This motivated \citet{85BoSexx.CS} to perform
\ai\ calculations of the dipole moment for CS over a range of internuclear
separations (2.5 $a_0  < R < 3.3 a_0$).  The resulting DMC yielded results
that were in good agreement with the Stark values and so it was extended to
longer and shorter internuclear distances by \citet{87PiTiCh.CS} using a
Pad\'{e} approximant:
\begin{equation}
M(x) = \frac{N_{0}+N_{1}x}{1+D_{1}x+D_{2}x^{2}+D_{3}x^{3}+D_{4}x^{4}+D_{5}x^{5}}
\end{equation}
\noindent where $x$ = $(R-R_{\rm e})/R_{\rm e}$. This function and
coefficients $N_{i}$, $D_{i}$ were used to generate turning points for input
to LEVEL. The results of \citet{87PiTiCh.CS} were reproduced  to the
precision quoted in their paper by this approach.

\begin{figure}
\begin{center}
\scalebox{0.3}{\includegraphics{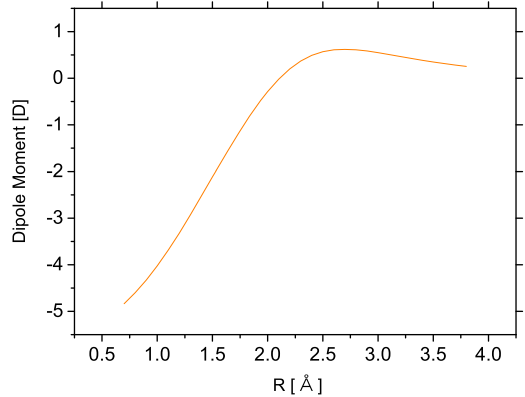}}
\caption{Dipole moment curve for CS obtained from the Pad\'{e} expansion parameters of \citet{87PiTiCh.CS}.}
\label{fig:dms}
\end{center}
\end{figure}

\subsection{Partition Functions}

Partition function values for all 8 isotopologues of CS were
calculated by a direct sum of all calculated energies for a range of
temperatures. We determined that our partition function is at least
95\% converged at 3000 K and much better than this at lower
temperatures. Therefore temperatures up to 3000 K were considered.
Partition function values for the parent isotopologue from CDMS and
HITRAN \citep{11LaGaLa.partfunc} are compared to the present work in
Table~\ref{tab:pf}.

At lower temperatures the CDMS partition function values are expected
to be the most accurate and our values compare very well in this case.
At higher temperatures only partition function values from HITRAN are
available and our values are noticeably lower in this case. As the
HITRAN partition function values are derived from using analytical
approximations rather than a direct sum over energy levels, and do not agree
as well with the CDMS values at lower temperatures, the results from
this work are expected to be more accurate.

For ease of use, we fitted our partition functions, Q, to a series expansion of the form used by \citet{jt263}:
\begin{equation}
\log_{10} Q(T) = \sum_{n=0}^6 a_n \left[\log_{10} T\right]^n \label{eq:pffit}
\end{equation}
with the $a_n$ values given in Table~\ref{tab:pffit}.

\begin{table}
\caption{Comparison of $^{12}$C$^{32}$S partition functions}
\label{tab:pf} \footnotesize
\begin{center}
\begin{tabular}{llll}
\hline
T(K) & This work & CDMS & HITRAN \\
\hline
150  & 127.9828 &  127.9818     & 128.1770 \\
300  & 256.3156 &  256.3136     & 257.0819  \\
500  & 437.5903 &  437.5868     & 439.6790 \\
1000 & 1018.57  & -             & 1026.38  \\
2000 & 2880.03  & -             & 2910.73  \\
3000 & 5705.80  & -             & 5803.37  \\
\hline
\end{tabular}
\end{center}
\end{table}

\begin{table}
\caption{Fitting parameters used to fit the partition functions. Fits are valid for temperatures between 500 and 3000 K.}
\label{tab:pffit} \footnotesize
\begin{center}
\begin{tabular}{lrrrrrrrr}
\hline
&  $^{12}$C$^{32}$S & $^{12}$C$^{33}$S &  $^{12}$C$^{34}$S   &  $^{12}$C$^{36}$S    &  {$^{13}$C$^{32}$S} & $^{13}$C$^{33}$S &  {$^{13}$C$^{34}$S} & $^{13}$C$^{36}$S \\
\hline
$a_0$ 	&	-80.005915	&	-76.155662	&	-80.005983	&	-80.004802	&	-79.895945	&	-79.450814	&	-79.753470	&	-72.151123	\\
$a_1$ 	&	135.900176	&	130.515908	&	135.900484	&	135.901969	&	136.196073	&	136.926482	&	136.119795	&	132.718091	\\
$a_2$ 	&	-92.069670	&	-88.464147	&	-92.068939	&	-92.067292	&	-92.226365	&	-92.982691	&	-92.125142	&	-99.251755	\\
$a_3$ 	&	32.390529	&	31.162671	&	32.390765	&	32.392690	&	32.426480	&	32.752272	&	32.286797	&	39.931597	\\
$a_4$	&	-6.167993	&	-5.950792	&	-6.168046	&	-6.170901	&	-6.170418	&	-6.233068	&	-6.095478	&	-9.136353	\\
$a_5$ 	&	0.599177	&	0.581747	&	0.599173	&	0.600214	&	0.599063	&	0.603143	&	0.582017	&	1.139224	\\
$a_6$ 	&	-0.022766	&	-0.022418	&	-0.022766	&	-0.022885	&	-0.022771	&	-0.022683	&	-0.021361	&	-0.060799	\\
\hline
\end{tabular}
\end{center}
\end{table}

\subsection{Line-List Calculations}

Line lists were calculated for all stable isotopologues of CS
($^{12}$C$^{32}$S, $^{12}$C$^{33}$S, $^{12}$C$^{34}$S,
$^{12}$C$^{36}$S, $^{13}$C$^{32}$S, $^{13}$C$^{33}$S, $^{13}$C$^{34}$S
and $^{13}$C$^{36}$S). These line lists span frequencies of up to
50,000 \cm.  A summary of each line list is given in
Table~\ref{tab:finallevels}. All rotation-vibration states up to $v$ =
49 and $J$ = 258, and all transitions between these states satisfying
the dipole selection rule $\Delta J = \pm 1$, were considered.

\begin{table}
\caption{Summary of our line lists.}
\label{tab:finallevels}
\begin{center}
\begin{tabular}{lllllllll}
\hline
&  $^{12}$C$^{32}$S & $^{12}$C$^{33}$S &  $^{12}$C$^{34}$S   &  $^{12}$C$^{36}$S    &  {$^{13}$C$^{32}$S} & $^{13}$C$^{33}$S &  {$^{13}$C$^{34}$S} & $^{13}$C$^{36}$S \\
\hline
Maximum $v$ & 49 & 49 & 49 & 49 & 49 & 49 & 49 & 49 \\
Maximum $J$ & 258 & 258 & 258 & 258 & 258 & 258 & 258 & 258 \\
Number of lines & 548312 & 550244  & 554898 & 560733 & 577885 & 581375 & 584485 & 590320 \\
\hline
\end{tabular}
\end{center}
\end{table}

\section{Results}

The line lists contain over half a million transitions each. For
compactness and ease of use they are separated into energy state and
transitions files using the standard ExoMol format \citep{jt548},
which is based on a method originally developed for the BT2 line list
\citep{jt378}. Extracts from the start of the $^{12}$C$^{32}$S files
are given in Tables~\ref{tab:states} and \ref{tab:trans}. The full
line lists for all isotopologues considered can be downloaded from
CDS, via ftp://cdsarc.u-strasbg.fr/pub/cats/J/MNRAS/xxx/yy, or
http://cdsarc.u-strasbg.fr/viz-bin/qcat?J/MNRAS//xxx/yy or can be
obtained from www.exomol.com.

\begin{table}
\caption{Extract from start of states file for $^{12}$C$^{32}$S. The full file
can be downloaded from 
http://cdsarc.u-strasbg.fr/viz-bin/qcat?J/MNRAS//xxx/yy or 
www.exomol.com.}
\label{tab:states}
\begin{center}
\begin{tabular}{lrlll}
\hline
$n$ &   $\tilde{E}$      &  $g$  &  $J$ & $v$\\
\hline
1	&	0.000000	&	1	&	0	&	0	\\
2	&	1.634164	&	3	&	1	&	0	\\
3	&	4.902459	&	5	&	2	&	0	\\
4	&	9.804822	&	7	&	3	&	0	\\
5	&	16.341155	&	9	&	4	&	0	\\
6	&	24.511332	&	11	&	5	&	0	\\
\hline
\end{tabular}

\noindent
  $n$:   State counting number;\\
 $\tilde{E}$: State energy in \cm;\\
  $g$: State degeneracy;\\
$J$:   State rotational quantum number;\\
 $v$:   State vibrational quantum number.
\end{center}
\end{table}

\begin{table}
\caption{Extracts from the transitions file for$^{12}$C$^{32}$S.
The full file
can be downloaded from 
http://cdsarc.u-strasbg.fr/viz-bin/qcat?J/MNRAS//xxx/yy or 
www.exomol.com.}
\label{tab:trans}
\begin{center}
\begin{tabular}{lll}
\hline
       $F$  &  $I$ & $A_{FI}$\\
\hline
2	&	1	&	1.7471E-06	\\
3	&	2	&	1.6771E-05	\\
4	&	3	&	6.0640E-05	\\
5	&	4	&	1.4904E-04	\\
6	&	5	&	2.9766E-04	\\
7	&	6	&	5.2215E-04	\\
\hline
\end{tabular}

\noindent
 $F$: Upper state counting number;\\
$I$:   Lower state counting number;\\
$A_{FI}$:  Einstein A coefficient in s$^{-1}$.

\end{center}
\end{table}

Table~\ref{tab:studies} compares our CS line lists with the previous
ones from \citet{87PiTiCh.CS} and CDMS; note that the 2078 CS
transitions given in HITRAN-2012 \citep{jt557} are reproduced from
CDMS.  Although this is an assessment of the quantity of the data, not
its quality, it demonstrates the reason for computing the new line
lists, to provide a more comprehensive coverage of the problem.

\begin{figure}
\begin{center}
\scalebox{0.3}{\includegraphics{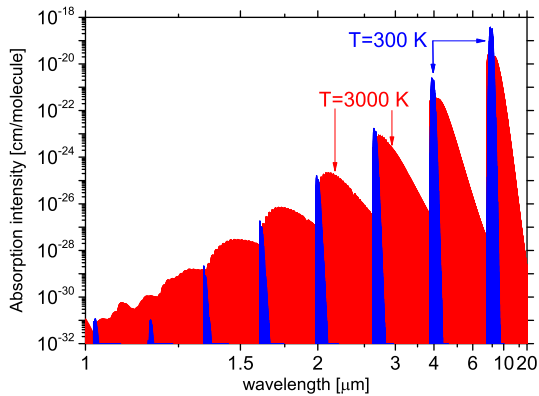}}
\caption{$^{12}$C$^{32}$S absorption spectrum at infrared wavelengths as calculated
in this work for a temperature of 300 K (narrow features) and 3000 K (broad features).}
\label{fig:overview}
\end{center}
\end{figure}

Figure~\ref{fig:overview} provides an overview of CS absorption at infrared wavelengths
as function of temperature. At 300 K, and below, the spectrum is dominated by a series
of vibrational bands starting with the fundamental band at about 8 $\mu$m. At higher temperatures
the bands become very much broader and their peak absorption is reduced.

%Illustrative comparisons of our calculated frequencies with observed line position %from the most
%recent infra-red study \citep{95RaBeDa.CS} are given in Table~\ref{tab:freq}. All %experimental
%frequencies are predicted within 0.02 \cm\ and often much better than this.

\begin{table}
\caption{Comparison of CS rotation-vibration line lists. Given are the number of
isotopologues considered,
the maximum values for vibrational ($v$) and rotational ($J$) states considered,
the maximum change in vibrational state ($\Delta v$) and whether intensity information
and a partition function are provided.}
\label{tab:studies}
\begin{center}
\begin{tabular}{lccc}
\hline
 Reference                   &\citet{87PiTiCh.CS}& CDMS & This work\\
\hline
Isotopes                     & 4              & 6 & 6   \\
maximum $v$                            &20                    & 4  & 49 \\
maximum $J$                            &200                   & 99   & 258 \\
maximum $\Delta v$& 4                     & 2  & 50 \\
Intensities? & Yes                   & Yes     & Yes \\
Partition
Function? & No                   & Yes      & Yes \\
\hline
\end{tabular}
\end{center}
\end{table}

\begin{figure}
\begin{center}
\scalebox{0.3}{\includegraphics{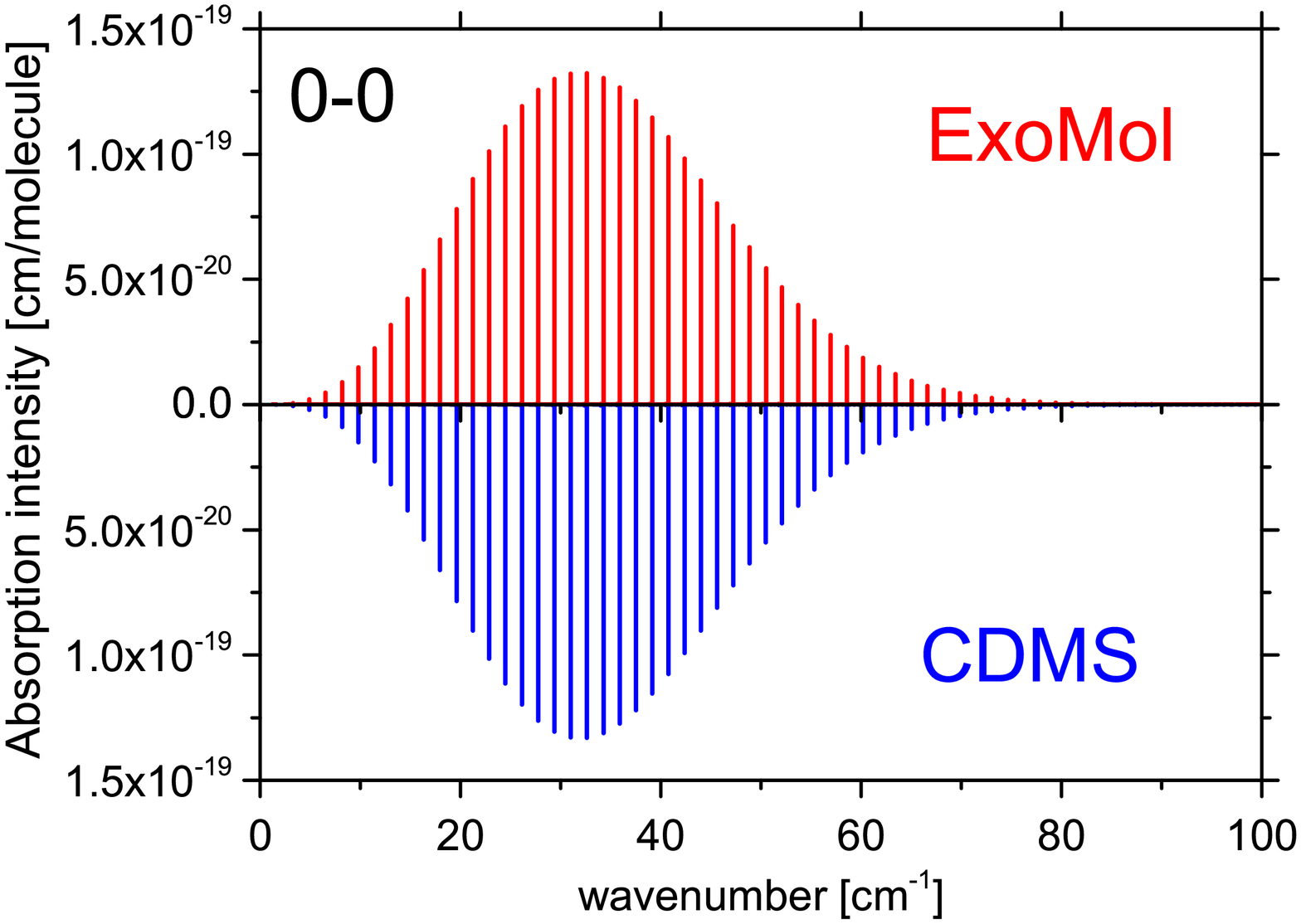}}
\scalebox{0.3}{\includegraphics{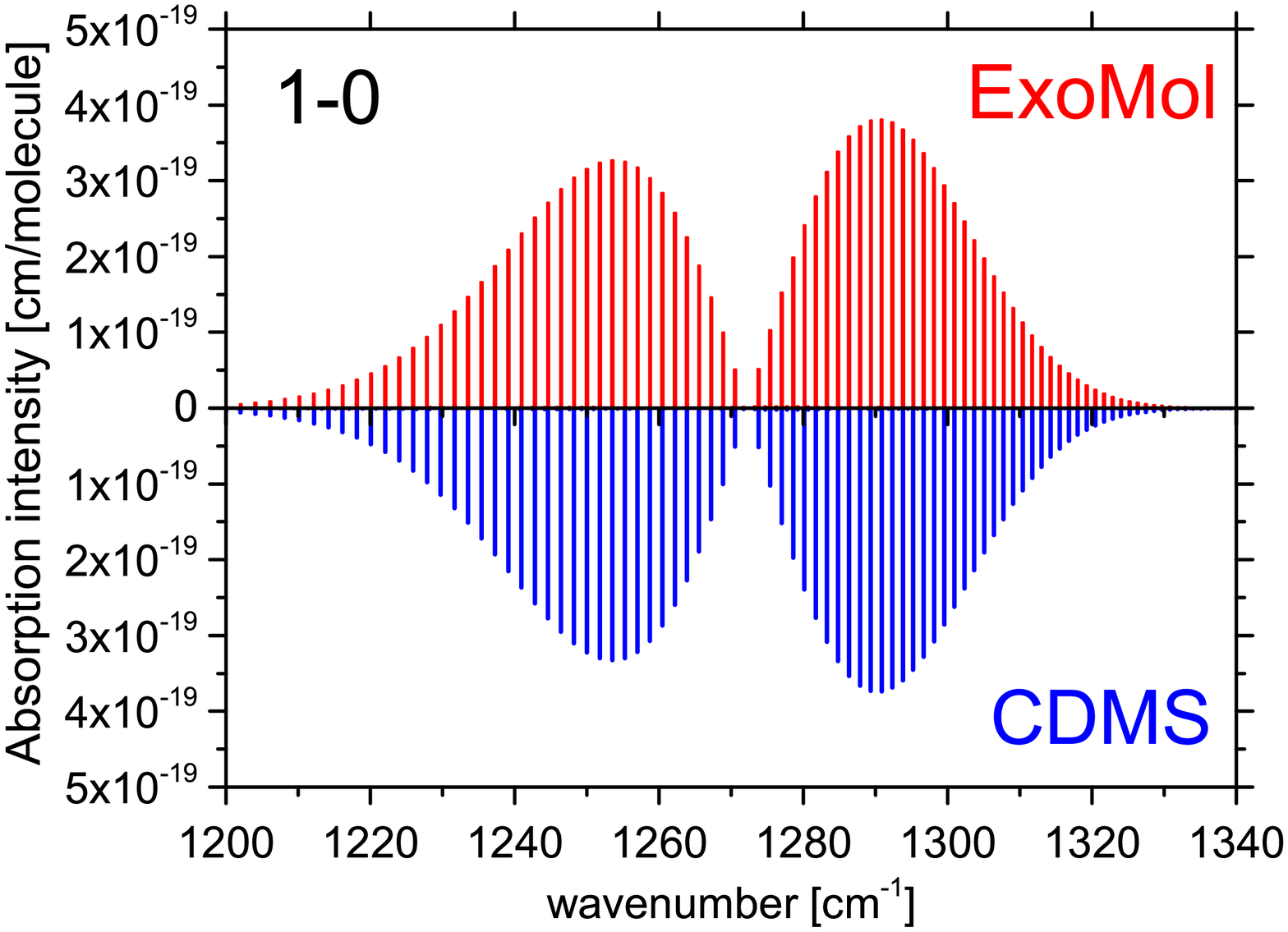}}
\scalebox{0.3}{\includegraphics{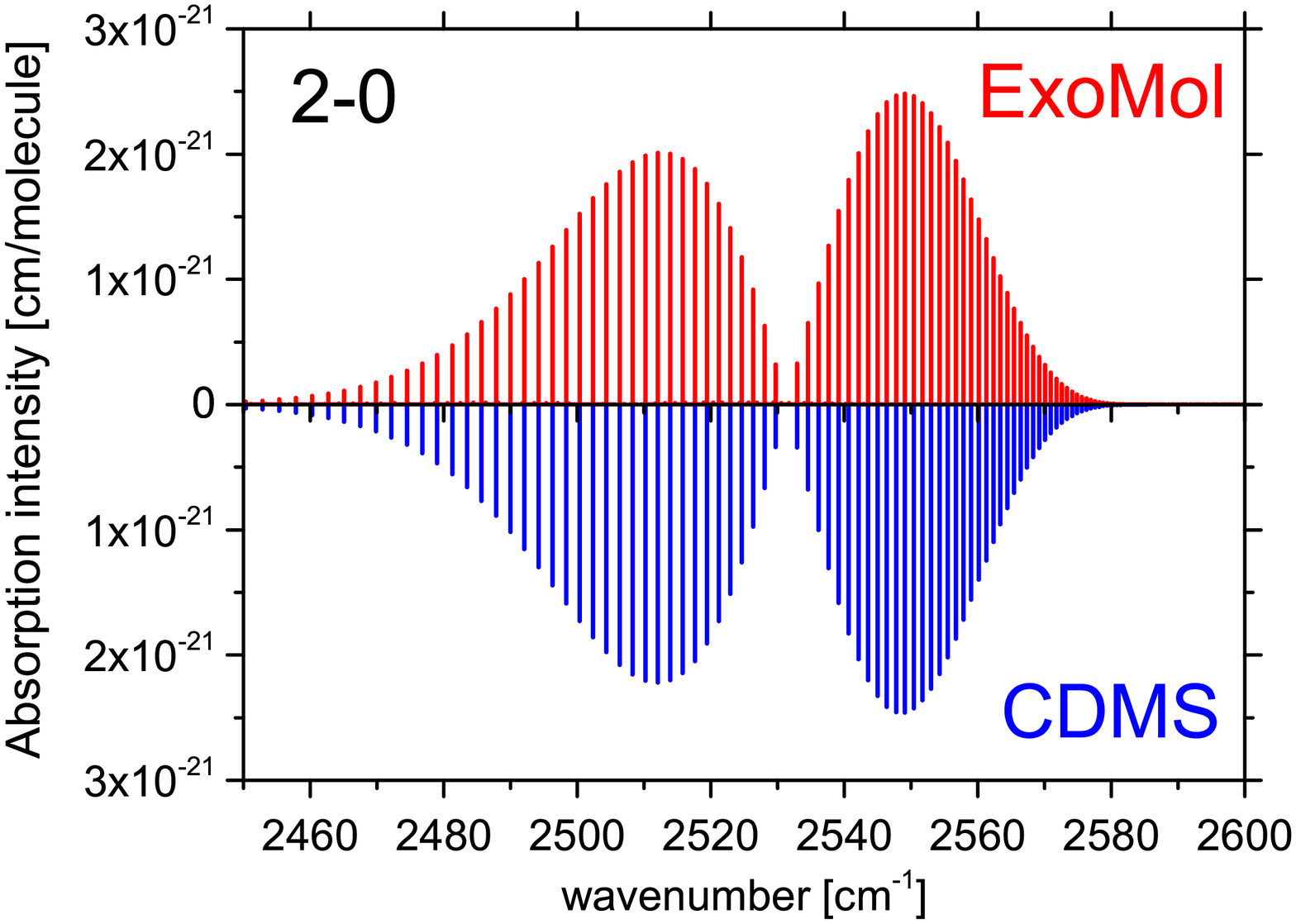}}
\caption{Absorption lines of $^{12}$C$^{32}$S at 300 K: ExoMol verses CDMS}
\label{fig:cdmslines}
\end{center}
\end{figure}

Comparisons with the CDMS rotational, $v\p - v\pp$ = 1 -- 0 and $v\p - v\pp$
= 2 -- 0 lines for $^{12}$C$^{32}$S are presented
in Figure~\ref{fig:cdmslines}. The agreement is excellent for both frequency
and intensity.

\begin{figure}
\begin{center}
\scalebox{0.3}{\includegraphics{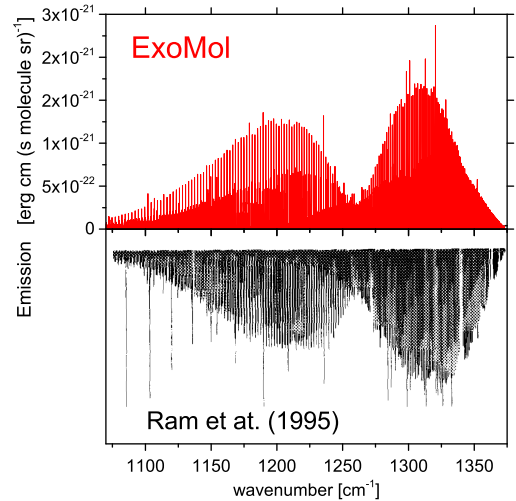}}
\caption{$^{12}$C$^{32}$S at 2300 C: Emission spectrum upper, Ram et al. (1995); Emission lines lower, ExoMol. [Reprinted from Ram et al. (1995)]}
\label{fig:ram1}
\end{center}
\end{figure}

\begin{figure}
\begin{center}
\scalebox{0.3}{\includegraphics{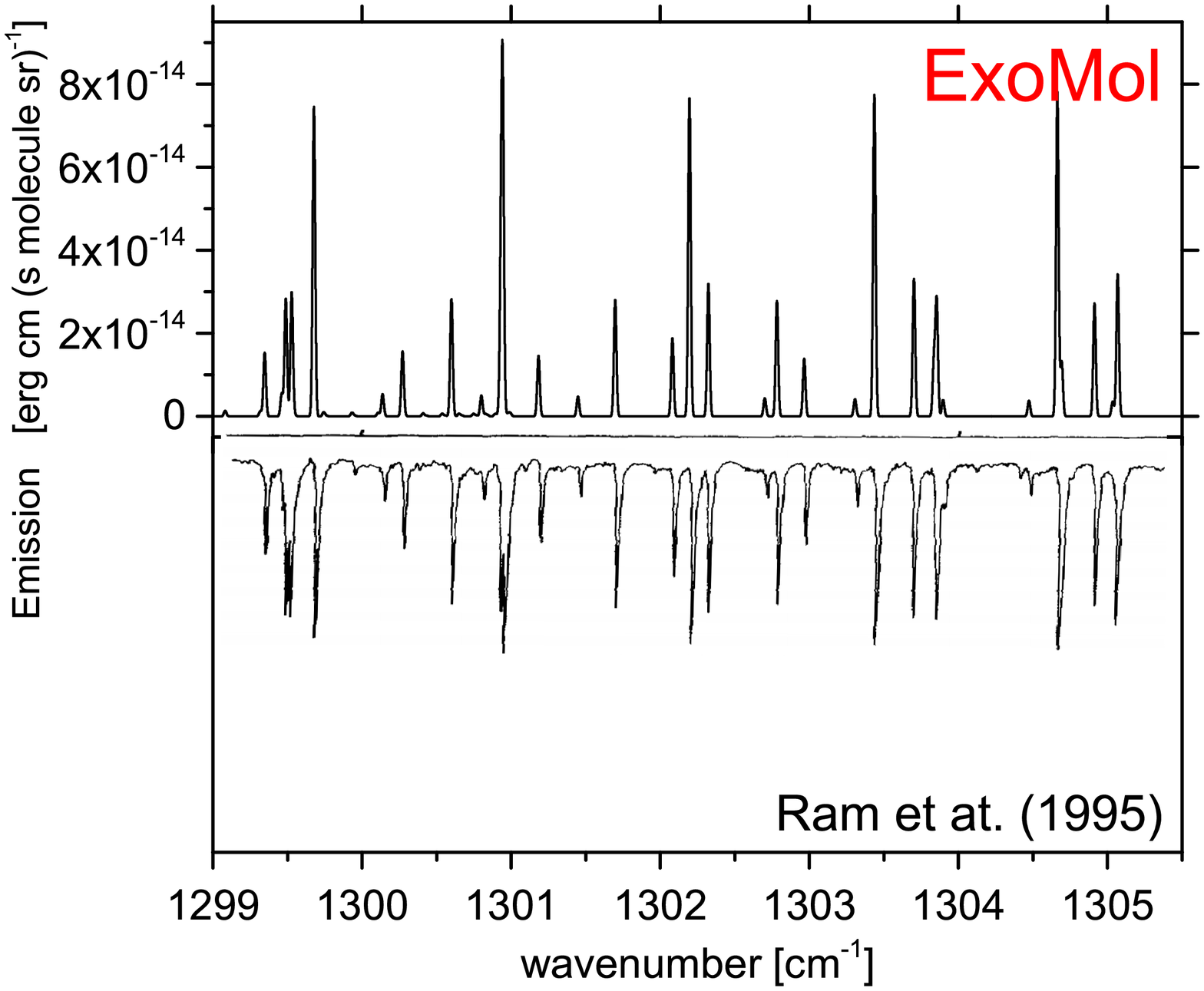}}
\caption{Emission spectrum of $^{12}$C$^{32}$S at 2300 C: upper, Ram et al. (1995); lower, ExoMol. [Reprinted from Ram et al. (1995)]}
\label{fig:ram2}
\end{center}
\end{figure}

\citet{95RaBeDa.CS} give two figures showing their observed spectrum,
a compressed view of the vibration-rotation bands (1000 - 1400 \cm) and
a portion of the R-branch region (1290 - 1310 \cm). Emission cross-sections 
for $^{12}$C$^{32}$S were simulated using a
Gaussian line shape profile with HWHM = 0.01 \cm\ as described in
\citet{jt542}. The resulting synthetic emission spectra are compared
to the experimental spectra in Figures~\ref{fig:ram1} and ~\ref{fig:ram2}.
For the former we the band structure and intensity ratio in our
calculated spectra is very similar to the experiment. For the latter there is
generally good agreement; however the intensities
of five strongest lines are almost 50~\%\ larger  in the theoretical spectrum
which may be due to saturation effects in the measured spectrum.

%to acccurately reproduce an experimental fourier transform spectrum one would have to take into consideration
%instrument line shape, path length, gas concentration and the quantum number dependence of line widths.

\section{Conclusions}

In the present work we have computed comprehensive line lists for all stable
isotopologues of carbon monosulphide. We determined a PEC using LEVEL and
modified potential parameters from the literature. We then substituted
calculated energies in the states file with energies derived directly from
experimental frequencies to match the experimental accuracy. This accuracy
should extend to all predicted transition frequencies up to at least $v$ = 9
and $J$ = 106 for $^{12}$C$^{32}$S, and $v$ = 5 and $J$ = 80 for $^{13}$C$^{32}$S,
the experimental ranges. Based on comparisons with other experiments the frequencies
for the remaining isotopologues should be predicted to sub-wavenumber accuracy
at least for v < 3 and J < 21. Einstein~A coefficients were computed
from a dipole moment curve taken from the literature. Comparisons with the
semi-empirical CDMS database suggest that the pure rotational, $v\p-v\pp$ = 1
-- 0, and $v\p-v\pp$ = 2 -- 0 intensities are accurate.

The results are line lists for rotation-vibration transitions within the
ground states of $^{12}$C$^{32}$S, $^{12}$C$^{33}$S, $^{12}$C$^{34}$S,
$^{12}$C$^{36}$S, $^{13}$C$^{32}$S, $^{13}$C$^{33}$S, $^{13}$C$^{34}$S and
$^{13}$C$^{36}$S, which should be accurate for a range of temperatures up to
at least 3000~K. The line lists can be downloaded from CDS or from
www.exomol.com.

Finally we note that, although our line lists are more comprehensive,
for the purposes of high resolution radio astronomy and far-infrared studies of
the low temperature objects, the CDMS line lists are recommended.

\section*{Acknowledgements}

This work is supported by ERC Advanced Investigator Project 267219.

\bibliographystyle{mn2e}
%\bibliography{journals_astro,CS,jtj,methods,diatomic,exogen,partition}

\label{lastpage}

\end{document}